\documentclass[preprint,12pt]{elsarticle}

\usepackage{graphicx}
\usepackage{amssymb}
\usepackage{lineno}
\usepackage{lipsum}
\usepackage{amsmath}

\makeatletter
\def\ps@pprintTitle{%
 \let\@oddhead\@empty
 \let\@evenhead\@empty
 \def\@oddfoot{}%
 \let\@evenfoot\@oddfoot}
\makeatother

\begin{document}

\begin{frontmatter}


\title{The Distributed Bloom Filter}

\author{Lum Ramabaja}
\ead{lum@bloomlab.io}

\author{Arber Avdullahu}
\ead{arber@bloomlab.io}

\address{Bloom Lab}




\begin{abstract}
The Distributed Bloom Filter is a space-efficient, probabilistic data structure designed to perform more efficient set reconciliations in distributed systems. It guarantees eventual consistency of states between nodes in a system, while still keeping bloom filter sizes as compact as possible. The eventuality can be tweaked as desired, by tweaking the distributed bloom filter's parameters. The scalability, as well as accuracy of the data structure is made possible by combining two novel ideas: The first idea introduces a new, computationally inexpensive way for populating bloom filters, making it possible to quickly compute new bloom filters when interacting with peers. The second idea introduces the concept of unique bloom filter mappings between peers. By applying these two simple ideas, one can achieve incredibly bandwidth-efficient set reconciliation in networks. Instead of trying to minimize the false positive rate of a single bloom filter, we use the unique bloom filter mappings to increase the probability for an element to propagate through a network. 

We compare the standard bloom filter with the distributed bloom filter and show that even with a false positive rate of 50\%, i.e. even with a very small bloom filter size, the distributed bloom filter still manages to reach complete set reconciliation across the network in a highly space-efficient, as well as time-efficient way.

\end{abstract}

\begin{keyword}
Bloom Filter \sep Distributed Bloom Filter \sep Distributed Systems
\end{keyword}

\end{frontmatter}

\section{Introduction}
\label{S:1}

\subsection{The Bloom Filter}
\label{S:bf}
Bloom filters are  space-efficient probabilistic data structures that can be used in all sorts of problems. They have been successfully used for web caching \cite{LiFan2000SummaryProtocol}, for fully decentralized computations of aggregate functions \cite{Pournaras2013ANetworks}, for memory-efficient genome assembly \cite{Melsted2011EfficientFilter}, for set reconciliation in distributed systems \cite{Guo2013SetFilters}, for more efficient block propagation in blockchain architectures \cite{PinarOzisik2017Graphene:Reconciliation}, and many other things. 

Even though there are many variants of bloom filters, most of them still have the same core idea: Bloom filters can quickly verify if an item is present in a set with minimal space requirements. This is achieved by sacrificing some precision for less space. When checking for the presence of an element in a bloom filter for example, one can get some false positive matches, but never false negative matches. In other words, a bloom filter can either prove that an item is not in a set, or that it might be in a set. One could use a hash table to achieve the same thing without the probabilistic nature of a bloom filter, but the minimal space requirements bloom filters offer, make it a useful data structure for many problems. 

\begin{figure}[h]
\centering\includegraphics[width=0.6\linewidth]{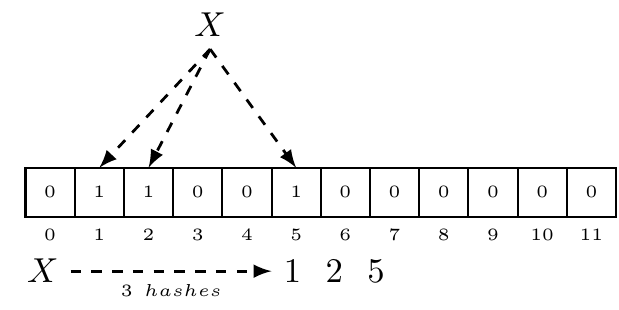}
\caption{Inserting an element into a bloom filter.}
\label{fig:bf}
\end{figure}

One big difference to most other data structures when inserting elements to a bloom filter, is that bloom filters do not actually store the elements themselves. That is in fact why bloom filters are so space-efficient. Instead of storing the actual elements, we perform a cleverly designed mapping. To better understand how this works, let's go over the algorithm:
\begin{enumerate}
\item \textit{k} independent hash functions are defined (where ''\textit{k}'' is the number of hash functions used).
\item An \textit{m} bit long zero bit array is defined.
\item When inserting elements to a bloom filter, we hash the element \textit{k} times, as defined in the first step. Each hash value is used to point to an index of the zero bit array (defined at step two). The bits at those indices are then switched from zero to one, as depicted in figure \ref{fig:bf}.
\end{enumerate}

Let's say that we inserted a few elements to a bloom filter, and now we want to check if a certain element is in it. To do so, we simply hash the element $k$ times and lookup the given indices. If the bit of one of the \textit{k} indices in the bloom filter is zero, we can conclude that the given element was never inserted into the bloom filter. In other words, we can always know if an element is not in the set, i.e. false negatives do not occur in the bloom filter. Things are a bit different when we want to check if an element \textit{is} present in the set. This is the kind of query that can result in a false positive.

\begin{figure}[h]
\centering\includegraphics[width=0.6\linewidth]{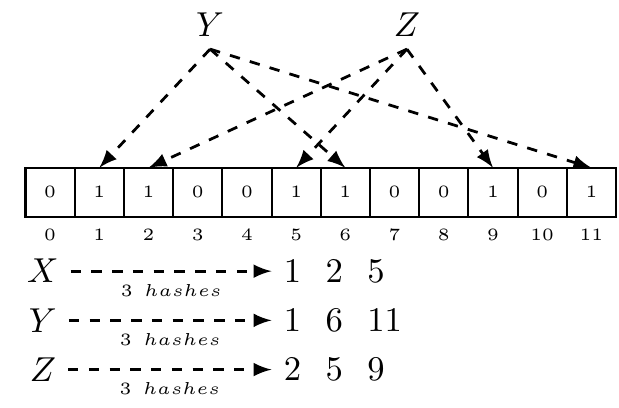}
\caption{ Inserting multiple elements into a bloom filter. ''\textit{X}'' was never inserted, yet it appears as if it has, i.e. a false positive has occurred.}
\label{fig:insertingBf}
\end{figure}

To better understand why this is the case, let's look at figure \ref{fig:insertingBf}.
We have element $Y$ and $Z$ and three hash functions to map values to the bloom filter. Let's assume that the resulting hashes for $Y$ are $\{11, 6, 1\}$ and those for $Z$ are $\{5,2,9\}$. We then switch all the bits of the given indices to one. If someone wants to verify if $Y$ or $Z$ are truly present in the set, one would get a positive result, this is because all the indices for $Y$ and $Z$ are one. There is however a problem with this. If there is another element, let's call it $X$, which happens to have $\{1,5,2\}$ as its hashes, one will get a false positive. The bits at $\{1,5,2\}$ are already one, but we never inserted $X$ to the bloom filter. As a result there is no way for us to know if $X$ really was inserted into the bloom filter or not.

Fortunately one can lower the chances for false positives by tweaking the parameters of the bloom filter: \textit{m} (the bloom filter size), \textit{n} (the number of elements inserted into the bloom filter), and \textit{k} (the number of hash functions used for the bloom filter). By knowing these values, we can compute the false positive rate of a bloom filter as: 

\begin{equation}
\label{eq:bf_fpr}
\left(1-\left(1-\frac{1}{m}\right)^{kn}\right)^k
\end{equation}

Let's unpack this formula: $1 - \frac{1}{m}$  shows the probability that a given bit in the bloom filter is not switched to one by a hash function.  $\left(1- \frac{1}{m}\right)^k$ shows the probability that a given bit is still zero, given that $k$ hashes can potentially point to it. And $\left(1- \frac{1}{m}\right)^{kn}$ is the probability that a particular bit is still zero, after $n$ elements were inserted. When we check for the presence of an element however, we do not look only at the value of one index, but at $k$. This is why our final formula for the false positive rate is $\left(1-\left(1-\frac{1}{m}\right)^{kn}\right)^k$, which gives the probability that $k$ indices will be 1 after inserting $n$ elements. Thus by choosing a proper $m$ and $k$ for a given $n$, one can design bloom filters with predetermined false positive rates.

Another important fact to consider, is that every hash function used has to be independent and its mappings uniformly distributed, otherwise we end up with non-random mappings in the bloom filter. The larger the bloom filter gets, the more hash functions we have to use. This can become a problem, as hash functions are computationally quite expensive and can slow down bloom filters. Because of this, people very often use faster non cryptographic hash functions, such as the Murmur non-cryptographic hash function to populate bloom filters. Distributed Bloom Filters on the other hand do store element hashes, but do not use hash functions for every element when populating bloom filters. This makes distributed bloom filters significantly faster to compute.

\subsection{Set reconciliation via bloom filters }
\label{S:set_reconciliation}
Set reconciliation is a problem in computer science where two nodes $A$ and $B$, each holding a set of elements $S_A$ and $S_B$, try to find the set difference $D_{A-B}= S_A - S_B$ and $D_{B-A}= S_B - S_A$ with as minimal traffic as possible. In other words, each node needs to figure out which elements of the other node it does not have, without having to send, or ask for the whole set of elements. Especially in peer-to-peer networks, if a node already has a subset of the elements from another node, it is desirable to ask somehow only for the missing elements. 

\begin{figure}[h]
\centering\includegraphics[width=0.5\linewidth]{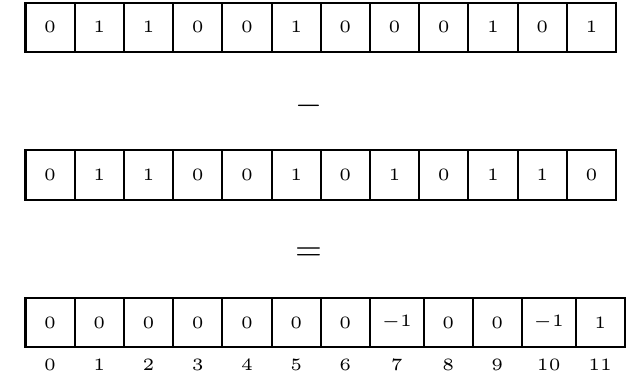}
\caption{Reducing a bloom filter from another bloom filter}
\label{fig:rbf}
\end{figure}

Nodes in a distributed system often have to synchronize states with one-another. Bloom filters are a great way for achieving such set reconciliation, as they require minimal space for transmission. Instead of sending a whole set to another node for set reconciliation, one can send a very small bloom filter. The node that receives the bloom filter can then create a bloom filter with its own elements, and reduce its own bloom filter with the one it received from its peer,  as shown in figure \ref{fig:rbf}. If any of the new values are -1, or in the negative range in general (in the case of counting bloom filters), then the node knows that the other node has some elements that it does not posses. It can then ask for the element(s) that point to the given indices. If any of the values on the other hand are in the positive range, the node knows that its peer is missing the elements whose hashes point at the indices of the positive values.

There is of course a catch with this approach - as mentioned earlier, there can be false positives. To guarantee a  very low false positive rate, one has to build a significantly larger bloom filter. So we can either sacrifice bloom filter size, which results in set inconsistencies across the network (which is not ideal for most tasks), or we use large bloom filters, which will lead to very few inconsistencies (which is still not desirable for some tasks), but increased bandwidth.  

This problem lead to the development of novel ideas such as the invertible bloom lookup tables (IBLT) \cite{Goodrich2011InvertibleTables}, or using counting bloom filters (CBF) for set reconciliation \cite{LiFan2000SummaryProtocol}. Both of these approaches work best when the set difference between the nodes is not too large. False positives however might still occur, as in any bloom filter data structure, especially when the set differences become larger than assumed.

We were interested to tackle this problem from another perspective. We wanted to have a data structure with the space advantages of standard bloom filters, but without the false positive drawback. 

\section{The Distributed Bloom Filter}
\label{S:distributed_bloom_filter}

\subsection{Unique Bloom Filter Mappings}
The Distributed Bloom Filter (DBF) is a probabilistic data structure meant for distributed systems that require quick synchronization in a space-efficient as well as time-efficient way. The way this is achieved is very simple: Instead of sending the same bloom filter to different nodes, each node sends a bloom filter with a unique mapping whenever it interacts with another node. When we do this, the probability for an element to replicate to another node slightly changes. Instead of calculating the false positive probability for an element in a bloom filter as in formula \ref{eq:bf_fpr}, we now calculate the probability for $N$ nodes not to distinguish the missing element:

\begin{equation}
    \label{eq:dbf1}
    \left[\left(1-\left(1-\frac{1}{m}\right)^{kn}\right)^{k}\right]^{N}
\end{equation}

$N$ in this case is the number of other nodes that are being contacted. This small change has some interesting implications. We are suddenly allowed to have a high false positive rate for our bloom filters, which means much smaller bloom filter sizes $m$ in relation to the size of the set $n$. As we are contacting each of the $N$ other nodes with a differently mapped bloom filter, the chances for all of the nodes to have a false positive hit for an element decreases with every added node. 

\begin{figure}[h]
\centering\includegraphics[width=0.7\linewidth]{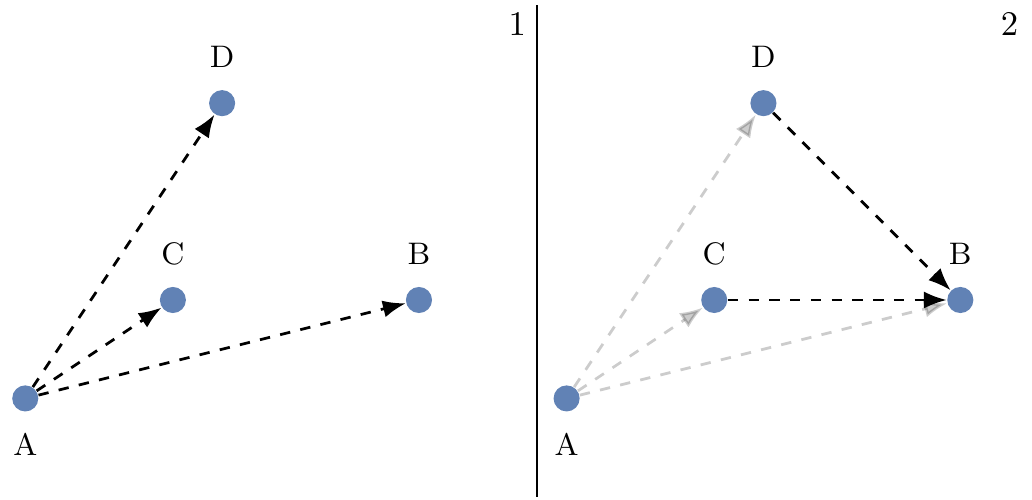}
\caption{A set of nodes contacting one another. In the left part, node A contacts B,C, and D, whereas in the right part nodes C and D contact node B}
\label{fig:transfering_DBFs}
\end{figure}

To better understand the implications of this, let's walk over a simple example depicted in figure \ref{fig:transfering_DBFs}: Let's say we have nodes $A$, $B$, $C$, and $D$, and let's say node $A$ has an element that the others do not have. In this scenario, node $A$ sends a unique bloom filter with a high false positive rate to each of the other nodes (as in the left part of figure \ref{fig:transfering_DBFs}). Two of its peers (node $C$ and $D$) figure out that they are missing an element, whereas node $B$ gets a false positive hit, i.e. it does not figure out that it is missing any elements. Now during the next iteration, if node $B$ gets contacted by node $A$, it will still have the same bloom filter mapping as before, which means it will have a false positive hit once again. But this time node $A$ is not the only one with the given element, $C$ and $D$ can also contact $B$. Since $B$ has a different bloom filter mapping with each node it interacts, the probability for it to distinguish the missing element(s) is now larger.

Notice how we are not interested for every node to identify that they are missing an element. All we want is a subset of the contacted nodes to identify the missing element(s). If the sent bloom filters each have a false positive rate of 50\% for example, 50\% of the contacted nodes will not have a false positive hit on average (for a certain element). This switches the perspective when it comes to the false positive rate. Instead of focusing on the false positive rate of a single bloom filter, we want to calculate the probability for an element to spread through a network. Once a subset of other nodes posses a given element, the chances for the element to further spread increases with each iteration. By using unique bloom filter mappings between nodes, one can thus replicate data across a network in a very bandwidth-efficient way.

Some readers might have noticed an obvious problem with the proposal so far: If we use a unique bloom filter for each node we interact with, we would have to hash all of the elements in our set again $k$ times, for every interaction we have. Hash functions are already timewise expensive enough, having to use them that often would not scale well.

\subsection{Element Insertions}
\label{sub:insertions}
To overcome this issue, we propose a new way to insert elements into bloom filters. We first assume that each node stores the cryptographic hashes of each of its elements. Most peer-to-peer systems require to compute element hashes anyways, so we can look at this as a step with no extra computation. 

\begin{figure}[h]
\centering\includegraphics[width=0.7\linewidth]{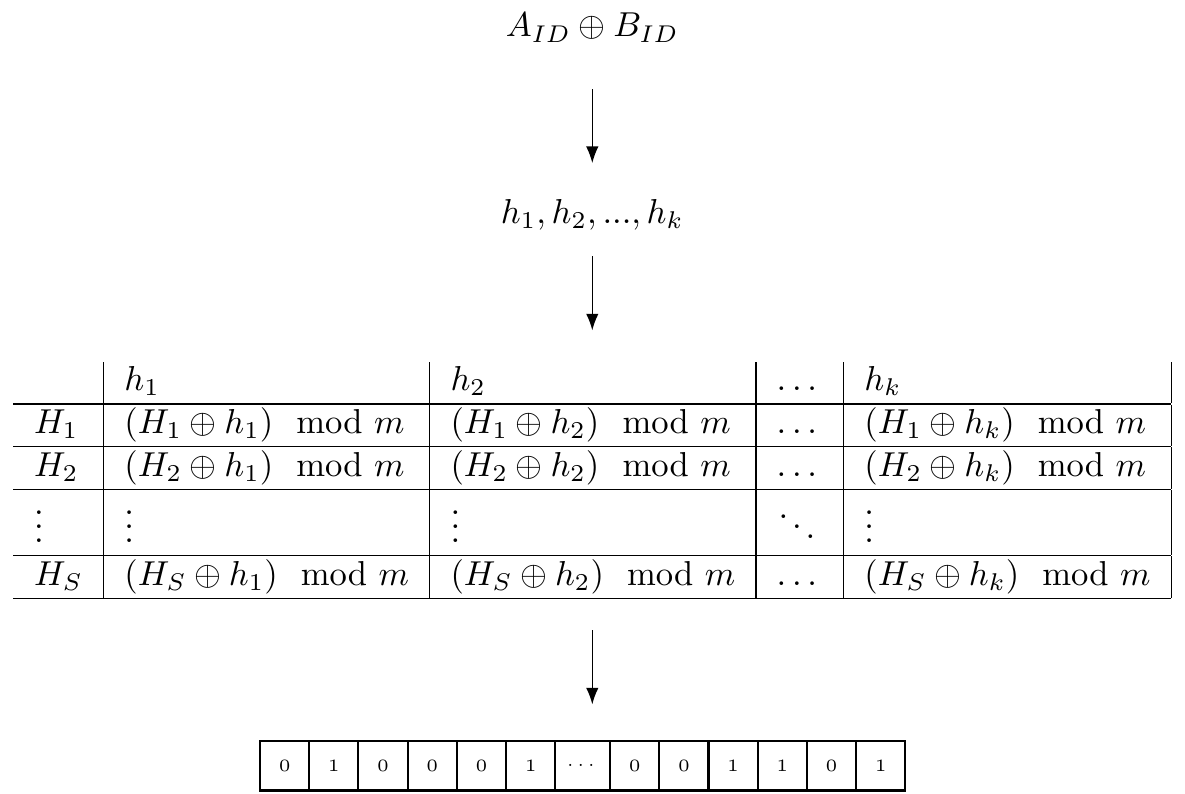}
\caption{First step: We choose a ''seed value'', in this case we compute the XOR of $A_{ID}$ and $B_{ID}$ (which represent the corresponding unique identifiers for a node pair). Second step: The XORed IDs get hashed $k$ times ($h_1$ to $h_k$ represent the generated hashes). Third step: Every element hash ($H_1$ up to $H_S$, where $S$ represents the set size) gets XORed with every hash generated in the second step. We then take modulo $m$ of each generated value (where $m$ is the chosen bloom filter size), so that every generated integer lies in the desired range. Step four: populate a bloom filter with the values of step three.}
\label{fig:dbf_algorithm}
\end{figure}

To generate quick, unique bloom filter mappings, we generate $k$ unique hashes $h$ between pairs of nodes using a ''seed value''. A seed value can be a nonce, a counter, a unique random value, or anything else. In this paper, the way we create the $k$ hashes is by XORing the two node identifiers, and hashing the resulting bit string $k$ times (as shown in step one and two of figure \ref{fig:dbf_algorithm}). The nature of the seed value thus can depend on the scenario.

Once we compute the $k$ hashes for a pair of nodes, We XOR each of the $k$ generated hashes with each element hash in our set. This will result in a $S{\times}k$ matrix of values, where $S$ represents our set size, and $k$ the number of generated hashes (as shown in step three of figure \ref{fig:dbf_algorithm}). We then apply the modulo operation with the desired bloom filter size to the matrix of values (so that the generated integers will be in the range of the bloom filter). Once this is done, we populate the bloom filter with the generated matrix values. 

This simple trick allows us to quickly generate a unique bloom filter with much fewer computational steps than in normal bloom filters. Since every element hash is random, and since all of the $k$ generated hashes are also random, the XOR of the values will also be random.

\subsection{Making the Distributed Bloom Filter More Efficient}
\label{Sub:cdbf}
We explained so far how a system that uses the DBF manages to create a unique bloom filter for each pair of nodes in an efficient way. The same pair of nodes however, in that setup, will have the same kind of mapping whenever they contact one-another. We also explained how a false positive hit in the case of the DBF depends on the number of peers a node contacts. Realistically speaking, that number cannot be too high in real-world applications. To achieve a guaranteed 100\% eventual set reconciliation between all interacting nodes, one can optionally use a different counter as a seed value during each interaction, or a nonce, for generating the $k$ hashes between a pair of nodes. That way, a pair of nodes can achieve a different bloom filter mapping each time they interact. That way we can have a guaranteed eventual data consistency across the network even when the nodes in the system have a small number of peers.

\section{Experiments}
\label{S:expt}

\subsection{Rationale} 
\label{Sb:rationale}
We are interested to know if the approach that the distributed bloom filter is more effective for set reconciliation in distributed systems, than that of the standard bloom filter. As the false positive rate of an individual distributed bloom filter does not reflect its set reconciliation abilities, we will need a better metric for a fair comparison. Since the end goal is to see which data structure guarantees better set reconciliation, our experiments will compare the performance of a network that uses the distributed bloom filter, with that of a network that uses the standard bloom filter. Both experiments will have all the same information. In other words, both systems will start with the same elements, same bloom filter parameters, and same neighbors. The only difference will be the kind of bloom filter used. For the distributed bloom filter experiments, we are going to use the element insertion proposal from subsection \ref{sub:insertions}, without the counter / nonce proposal of section \ref{Sub:cdbf}. At the end we will look at how many nodes will have converged to the same set of elements, and how many not. 

\subsection{Experimental Design}
\label{Sb:exp_design}
 
We will run two experiments on two networks to properly compare the performance of the DBF to the standard bloom filter. For both experiments, every node will have a randomly initialised subset of a set of elements. In other words, given a large set $S$, each node in the system will have a subset $T$, where each element in $T$ is randomly picked from the main set $S$. This means that each node in the system will have some elements of $S$ and possible overlaps with one-another. The goal of the experiments is to see if the nodes in the network can converge to $S$ by sending bloom filters to one-another. 
 
\subsubsection{First Experiment}
\label{sb:first_exp}
For the first experiment, we will have two networks with identical nodes, neighbors, elements per nodes, and bloom filter parameters. The only difference will be the kind of bloom filter used. In this experiment, we will both have a fixed bloom filter size $m$, as well as a fixed upper-bound for the false positive rate (which for this experiment will be50\%). If we look at formula \ref{eq:bf_fpr}, we can see that to have a fixed $m$ and a fixed false positive rate, one also needs a fixed $n$. In this experiment, instead of nodes choosing $m$ according to their set size $n$, they will choose $m$ according to the known upper bound $S$ and the system-wide false positive rate of 50\%. This means that for a node with fewer elements than $S$, the false positive rate will be smaller, but will approach the fixed false positive rate with every newly acquired element.
 
The design of the first experiment has several implications: First, since $m$ will be a fixed constant and every node knows the size of $S$, nodes can have one way communications (if we ignore the pinging part). Each node just computes their bloom filter, and sends it to other nodes. Second, since $m$ stays constant, the network with the standard bloom filter will always have the same kind of mapping. This gives the network with the standard bloom filter an advantage computationally-wise (nodes will not have to recompute their bloom filter from scratch each time), but the same network should logically also suffer performance-wise, because of the same bloom filter mappings. 
 
Both the network with the distributed bloom filter, as well as the one with the standard bloom filter will have 50 nodes. Each node has 10 randomly chosen neighbors, and will send bloom filters with a false positive rate of up to 50\%. The total set size of $S$ will be 1000, and each node will have 200 randomly selected elements from the main set. Each node sends periodically bloom filters to its neighbors. At the end we want to see how many nodes from the networks reached a set of size 1000 (in the distributed bloom filter network, as well as the standard bloom filter network).
 
\subsubsection{Second Experiment}
\label{sb:second_exp}
The second experiment will also have two networks with identical nodes, neighbors, elements per nodes, and bloom filter parameters (all parameters are in fact identical to the first experiment). The difference in this experiment, is that we do not assume an upper bound to the number of elements to compute the bloom filters. This means that we are going to send bloom filters with a fixed false positive rate (which in this experiment is again 50\%), but a changing $m$. This is because in this experiment $m$ will be chosen based on the nodes actual $n$, instead of a system-wide constant $S$ like in the first experiment. The nodes in this experiment will also send bloom filters with a false positive rate of 50\%, but their bloom filter sizes will increase over time (with every newly acquired element). 
 
The design of the second experiment has also some interesting implications: First, since $m$ will not be a fixed constant anymore, nodes will require to have two way communications (if we ignore pinging) to ensure a fixed false positive rate. The way this is done is straight forward, nodes first have to send their set size to one another, then pick the larger set size between the two sets, and then use that as the parameter for the bloom filter parameter selection.
 
We do this, because we want to make sure that the bloom filter parameters $m$ and $k$ are chosen based on the greater $n$ from the two nodes. If we would choose $min(n_A, n_B)$ instead (where $n_A$ is the set size of one node, and $n_B$ is the set size of the other node), one node would end up with a larger false positive rate than intended (i.e. one node would have a larger false positive rate than the allowed upper bound). Let's take an example to better understand this: If $n_A = 50$ and $n_B = 100$, and we choose $n_A$ to calculate $m$ and $k$, instead of $n_b$ (the larger value), node $A$ will design a bloom filter with a size $m$ that reflects its $n_A$. To find any missing values, node $B$ now also has to compute a bloom filter with the same $m$ and $k$. $B$ however has a much larger set size than $A$, which means $B$'s bloom filter will have a much larger false positive rate. We do not want that to happen. Thus picking the larger set size solves this issue.
 
The second interesting implication is that since $m$ will not be constant anymore, the nodes in the network with the standard bloom filter will always have a new bloom filter mapping. This gives the network with the standard bloom filter the same advantage that the distributed bloom filter has (i.e. unique mappings), but it will have to recompute new bloom filters every iteration, which is significantly more expensive for the standard bloom filter.

Both the network with the distributed bloom filter, as well as the one with the standard bloom filter will have 50 nodes. Each node has 10 randomly chosen neighbors, and will send bloom filters with a false positive rate of 50\%. The total set size of $S$ will be 1000, and each node will have 200 randomly selected elements from the main set. Each node sends periodically bloom filters to its neighbors. At the end we want to see how many nodes from the networks reached a set of size 1000 (in the distributed bloom filter network, as well as the standard bloom filter network).

\newpage
\subsection{Results}

\subsubsection{First Experiment}

\begin{figure}[h]
\centering\includegraphics[width=0.55\linewidth]{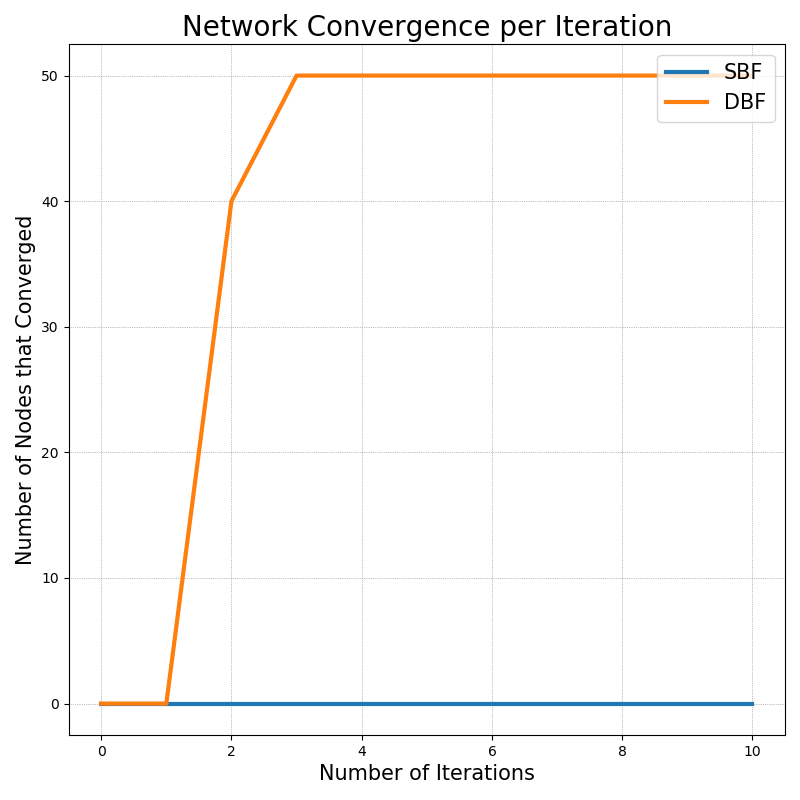}
\caption{A graph showing the rate of convergence for the distributed bloom filter (DBF) and standard bloom filter (SBF) with a fixed $m$. Both networks used bloom filters with a false positive rate of up to  50\%. The x-axis represents the number of iterations (each node sent bloom filters to neighbors during an iteration). The y-axis represents the number of nodes that converged to the total set $S$, i.e. the number of nodes that have reached the maximum number of elements (1000 in this case).}
\label{fig:exp1}
\end{figure}

Since we had a very high false positive rate for the experiments, it is expectable that none of the nodes in the network with the standard bloom filter converged to $S$. The maximum number of elements for this experiment was 1000. The median set size per node in the network with the standard bloom filter at the end was 793. Note how all the nodes in the network with the distributed bloom filter did converge to $S$. That is not always the case however, running the same experiment multiple times (but with differently initialized random node sets) might result in 1-3 nodes not converging completely. By using the slight modification proposed in section however \ref{Sub:cdbf}, one could always reach 100\% data consistency across the network.

\begin{figure}[h]
\centering\includegraphics[width=0.7\linewidth]{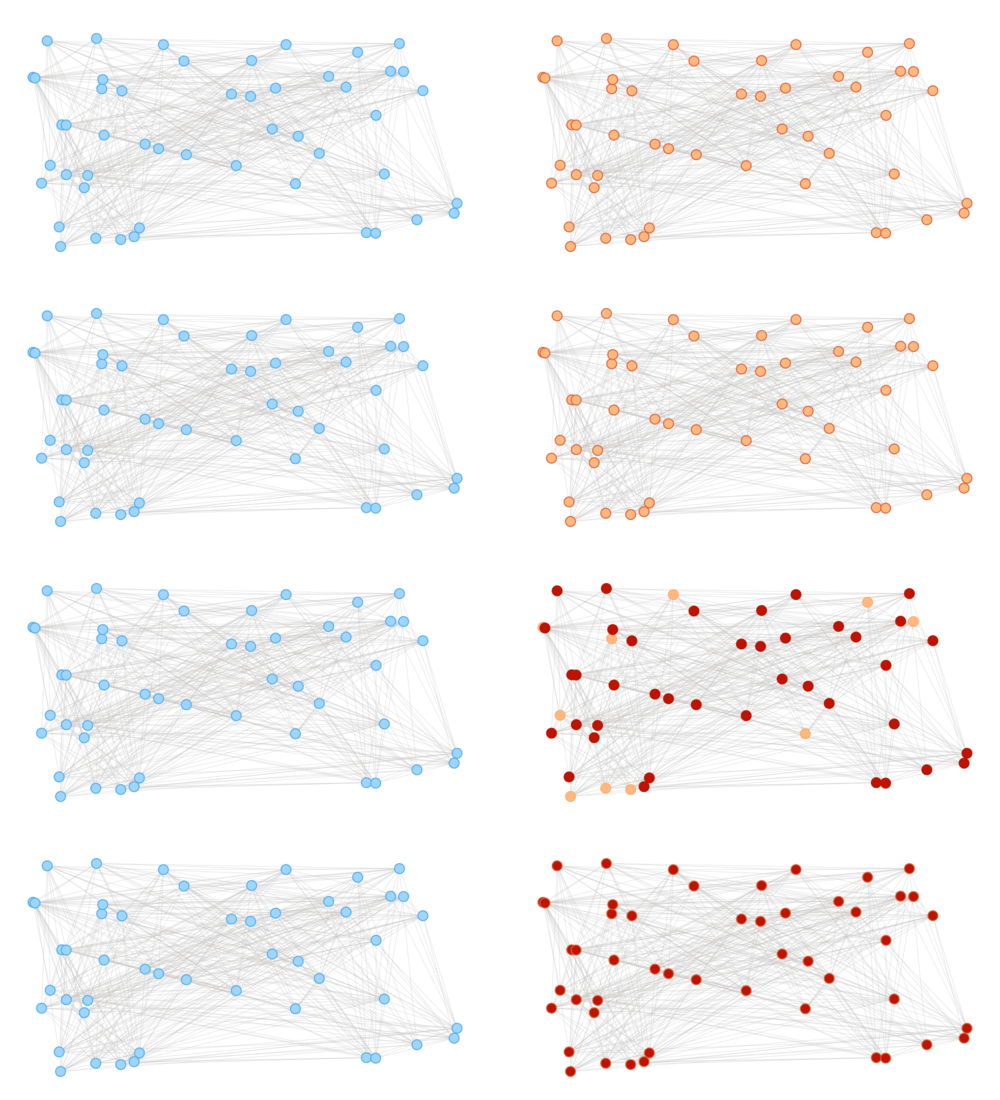}
\caption{The blue network (on the left column) used standard bloom filters, whereas the orange network (on the right column) used distributed bloom filters. Both networks had the same parameters as described in subsection \ref{sb:first_exp}. Each row represents an iteration. A node changing color means that it converged to the total set $S$.}
\label{fig:exp1}
\end{figure}

\newpage
\subsubsection{Second Experiment}

\begin{figure}[h]
\centering\includegraphics[width=0.6\linewidth]{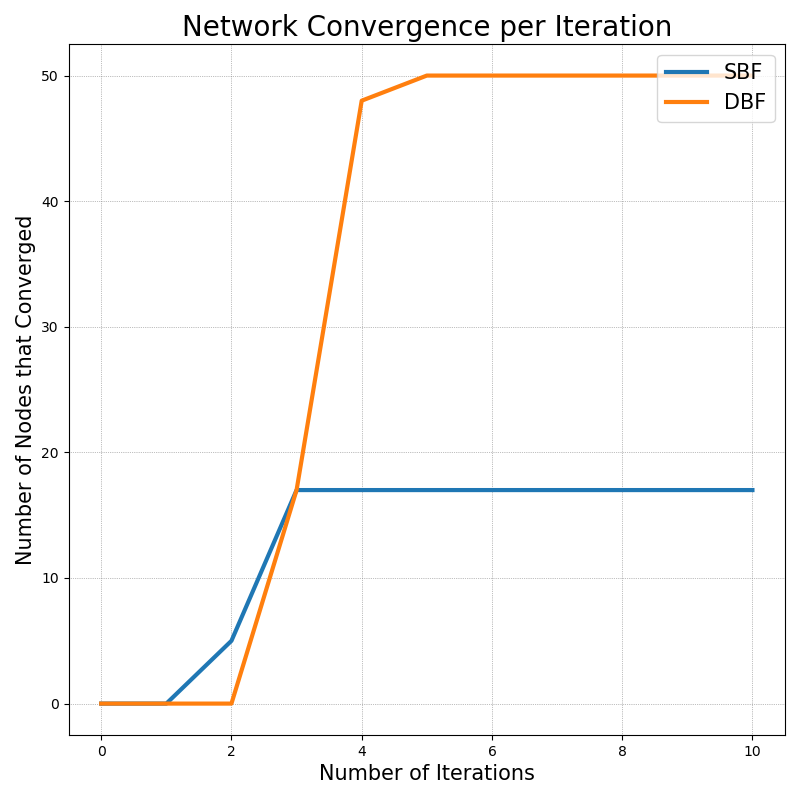}
\caption{A graph showing the rate of convergence for the distributed bloom filter (DBF) and standard bloom filter (SBF) without a fixed $m$. Both networks used bloom filters with a 50\% false positive rate. The x-axis represents the number of iterations (each node sent bloom filters to neighbors during an iteration). The y-axis represents the number of nodes that converged to the total set $S$, i.e. the number of nodes that have reached the maximum number of elements (1000 in this case).}
\label{fig:exp1}
\end{figure}

In this experiment, it is actually more interesting to analyze why the standard bloom filter performed as well as it did, as it might be an unexpected result for some readers (the false positive rate for each bloom filter was 50\% after all). The maximum number of elements for this experiment was again 1000. Remember however, only the false positive rate is fixed here, meaning $m$ changes whenever a node acquires a new element. This is in fact the reason why some nodes in the network with the standard bloom filter managed to get to the 1000 elements set size. Since $m$ changes whenever a node acquires a new element, the nodes in the network with the standard bloom filter will not be able to store their bloom filters anymore. Whenever $m$ changes, they have to recompute the whole bloom filter from scratch. As a result, the network with the standard bloom filter has to perform several fold more computations than the network with the distributed bloom filter. This is because the nodes using the standard bloom filter have to use hashes to populate their bloom filter, plus recompute their bloom filter each iteration. The median set size for the nodes that did not converge in the network with the standard bloom filter was 999. This high number (considering the high false positive rate) is due to the new bloom filters the network with the standard bloom filter had to compute in each iteration. Only 18 out of 50 nodes achieved to fully converge to $S$, however most nodes were very close to full convergence. In the case of the network with the distributed bloom filter on the other hand, all 50 nodes achieved to converge to $S$. The distributed bloom filter requires also several fold less computations in this experiment, than the network with the standard bloom filter.

\begin{figure}[!ht]
\centering\includegraphics[width=0.6\linewidth]{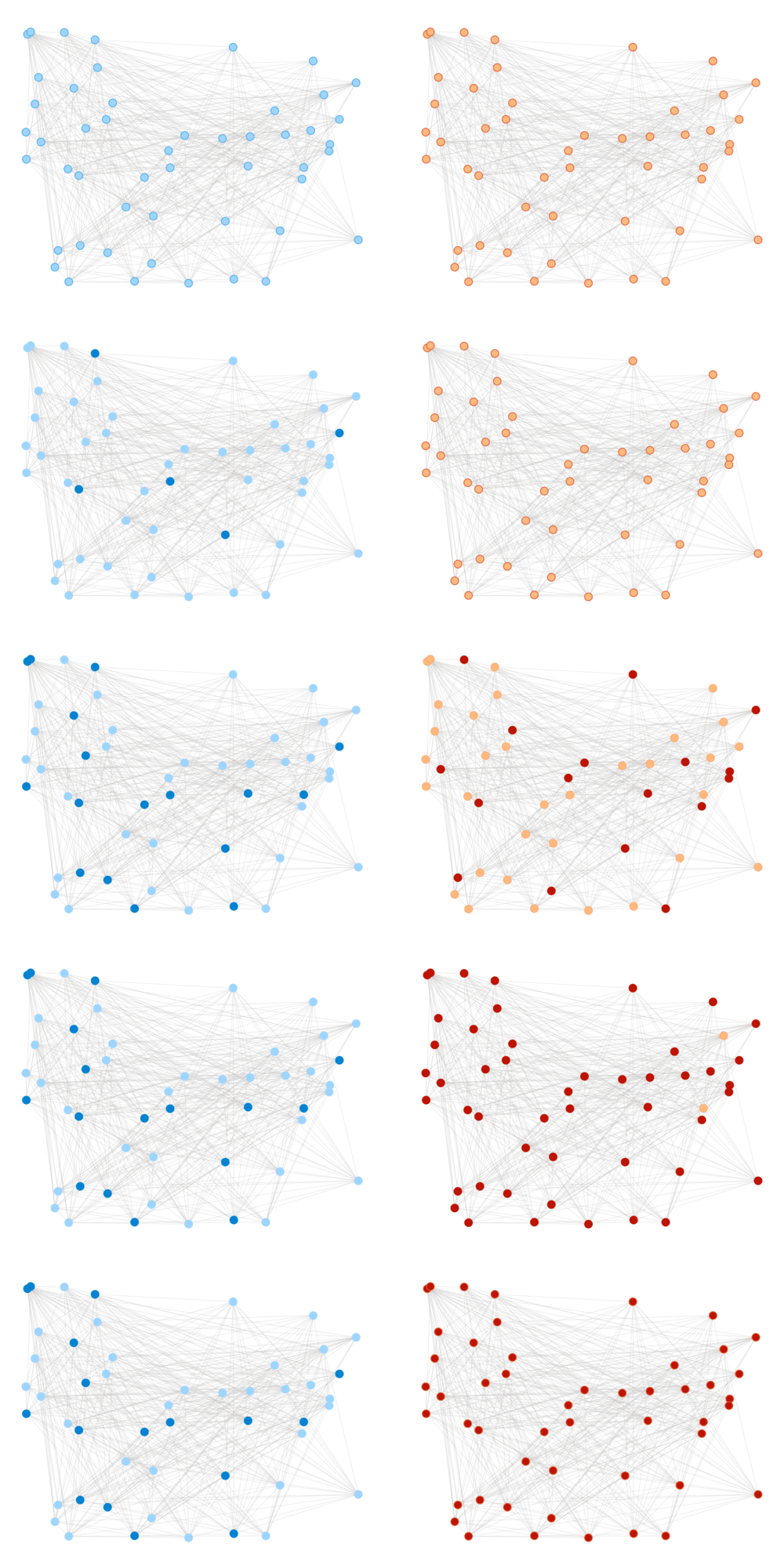}
\caption{The blue network (left column) used standard bloom filters, whereas the orange network (right column) used distributed bloom filters. Both networks had the same parameters as described in subsection \ref{sb:second_exp}. Each row represents an iteration. A node changing color means that it converged to the total set $S$.}
\label{fig:exp1}
\end{figure}

\newpage
\section{Conclusion}

We showed how by using unique bloom filter mappings, and a modified way to populate bloom filters, one can successfully achieve eventual data consistency across a whole network in a highly space-efficient, as well as time-efficient way. Based on the experimental results, we believe that the distributed bloom filter can be of interest for several applications: For more efficient caching in content distribution networks, mempool synchronization and block propagation in blockchain systems, distributed storage, peer-to-peer networks, and more.


\section{References}

\bibliographystyle{structure}
\bibliography{references.bib}
\end{document}